\documentstyle[12pt,epsfig]{article}
      \def\lsim{\raise0.3ex\hbox{$<$\kern-0.75em\raise-1.1ex\hbox{$\sim$}}}
\def\gsim{\raise0.3ex\hbox{$>$\kern-0.75em\raise-1.1ex\hbox{$\sim$}}}
\def\noi{\noindent}

\def\bea{\begin{eqnarray}}  \def\eea{\end{eqnarray}}
\def\beq{\begin{equation}}   \def\eeq{\end{equation}}

\def\beeq{\begin{eqnarray}} \def\eeeq{\end{eqnarray}}

\begin{document}
\begin{center}
\vbox to 1 truecm {}
{\Large \bf Anomalous suppression of $\pi^0$ production at large}\par
\vskip 3 truemm
   {\Large \bf transverse momentum in Au + Au and d + Au}\par \vskip 3 truemm
   {\Large \bf collisions at $\sqrt{{\bf s}_{\bf NN}} =$ 200 GeV} \\
\vskip 1.5 truecm
{\bf A. Capella$^{\rm a)}$, E. G. Ferreiro$^{\rm b)}$, A. B.
Kaidalov$^{\rm c)}$, D. Sousa$^{\rm b)}$}\\
\vskip 8 truemm

${\rm a)}$ Laboratoire de Physique Th\'eorique\footnote{Unit\'e Mixte de
Recherche UMR n$^{\circ}$ 8627 - CNRS}
\\ Universit\'e de Paris XI, B\^atiment 210,
F-91405 Orsay Cedex, France

\vskip 3 truemm

${\rm b)}$ Departamento de F{\'\i}sica de Part{\'\i}culas,
Universidad de Santiago de Compostela, 15782 Santiago de Compostela,
Spain

\vskip 3 truemm

${\rm c)}$ Institute of Theoretical and Experimental
Physics, 117259 Moscow, Russia

\vskip 3 truemm

\end{center}
\vskip 1 truecm
\begin{abstract}
We propose a model of suppression of large $p_T$-pions in heavy ion
collisions based on the interaction of the large $p_T$ pion with the
dense medium created in the collision. The model is practically the
same as the one previously introduced to describe $J/\psi$ 
suppression. Both the
$p_T$ and the centrality dependence of the data are reproduced. In
deuteron-gold collisions, the effect of the final state interaction
with the dense medium turns out to be negligibly small. Here the main
features of the data
are also reproduced both at mid and at forward rapidities.
\end{abstract}

\vskip 2 truecm
\noi LPT-Orsay-04-25\par
\noi February 2004
\newpage
\baselineskip = 20 pt
\pagestyle{plain}

\section{Introduction}
\hspace*{\parindent}
One of the most interesting results of the heavy ion program at RHIC is
the so-called jet quenching \cite{1r} \cite{2r}. The yield of
particles produced
in $AA$ collisions at mid-rapidities and large $p_T$ increases with
centrality much less than the number of binary collisions $n(b)$. For
most central collisions the large $p_T$ yield is suppressed by a factor
4-5 as compared to the result expected from this scaling. This
phenomenon is particularly interesting since it is not observed in
deuteron-gold
collisions at RHIC at mid-rapidities \cite{2r}. \par

The suppression of the yield with respect to the scaling in $n(b)$ is
well known in soft collisions. In this case the phenomenon is observed
in hadron-nucleus and nucleus-nucleus collisions at all energies. It is
well described in the framework of string models, such as the Dual
Parton Model (DPM) and the Quark Gluon String Model (QGSM), when
shadowing corrections are taken into account \cite{3r}. With increasing
$p_T$ the shadowing corrections decrease \cite{4r} and the scaling with
$n(b)$ is predicted in perturbative QCD. Actually, the observed
increase is even faster due to initial state interactions, the
so-called Cronin effect. \par

At very high energies the shadowing effects are very important for $hA$
and $AA$-collisions (see for example \cite{3r} \cite{4r}). These nonlinear
effects lead, as $s \to \infty$, to ``saturation'' of the distributions of
partons in the colliding hadrons and nuclei \cite{5r}. Detailed
calculations of shadowing effects at RHIC and LHC energies \cite{3r},
show that these effects are important for the description of inclusive
spectra, but the situation is still far from the ``saturation'' limit.
This is true for particles with average momentum transfer. For
particles with large momentum transfer, which we study in this paper, the
situation is different. It is well known that shadowing effects for
partons take place at very small $x$,
$x \ll x_{cr} = 1/m_N R_A$ where $m_N$ is the nucleon mass and $R_A$
is the radius of the nucleus. On the other hand, partons which produce
a state with transverse mass $m_T$ and a given value of Feynman $x_F$,
have $x = x_{\pm} = 1/2(\sqrt{x^2_F + 4m_T^2/s} \pm x_F)$. Thus, at fixed
initial energy $(s)$ the condition for existence of shadowing will not
be satisfied at large transverse momenta. For example in the central
rapidity region ($y^* =0$) at RHIC and for $p_T$ of jets (particles)
above 5(2) GeV/c the condition for shadowing is not satisfied and
these effects are absent. It was shown in ref. \cite{6r} that in this
region, where $x \ \gsim\ x_{cr}$, there are in general final state
interactions, which can be treated in a simple quasiclassical way.
These interactions lead, in particular, to an energy loss of a parton
(particle) in the dense medium produced in the collision. The situation
is very similar to production of heavy quarkonia in $pA$ and
$AA$ collisions \cite{7r}, where most of the present data correspond to
energies below the critical one and a simple probabilistic interpretation
can be applied. Note that these ``final state'' interactions are absent
in the shadowing region.\par

Hadrons loose a finite fraction of their longitudinal momentum due to
secondary interactions with hadrons of the nucleus. This is a
characteristic property of such theoretical models as DPM and QGSM and
agrees with approximate Feynman scaling of inclusive spectra in the
fragmentation regions. From this point of view, it is natural to expect
that a particle scattered at some non zero angle will also loose a
fraction of its transfer momentum due to final state interaction. This
is a characteristic property of soft hadronic interactions. In
perturbative QCD the situation is more complicated \cite{8r}. In the
following we will assume that final state interactions are mostly soft
ones. \par

The aim of the present work is to describe the suppression of the
yield of pions
in a framework, based on final state interactions, similar to the one
used by the authors in order to describe the suppression of $J/\psi$
\cite{9r}. In the latter case, the origin of the suppression is
twofold. On the one hand, the $c$-$\bar{c}$ pair interacts with
nucleons of the nucleus
(normal absorption or nuclear absorption, controlled by $\sigma_{abs}$).
On the other hand, the $c$-$\bar{c}$ pair (at times close to initial
time $\tau_0$) or the
$J/\psi$ (at larger times), interacts with the dense medium produced in
the collision (anomalous absorption, controlled by
$\widetilde{\sigma}$). In both cases, as a result of the interaction, a
$D\bar{D}$ pair is produced instead of a $J/\psi$. It turns out that in
hadron-nucleus collisions, the density of the medium is small enough
and such that the effect of the interaction with the medium is
negligible. Thus, this effect is only present in nucleus-nucleus
collisions -- hence its qualification of anomalous.\par

In the case of large $p_T$ production the particle does not disappear
as a result of the interaction but its $p_T$ is shifted to smaller
values. Due to the steepness of the $p_T$ distribution, the effect may
be quite large. Moreover, in this case there is also a gain of the 
yield at a given $p_T$ due to particles produced at larger
$p_T$ -- which have experienced a $p_T$ shift due to the interaction
with the medium. This gain is significantly smaller than the 
corresponding loss due to the steepness of the $p_T$ distribution.\par

Another difference with respect to the $J/\psi$ case is that here the
suppression vanishes at low $p_T$. Indeed, when the $p_T$ of the
produced particle is close to $<p_T>$, its $p_T$ can either increase
or decrease as a result of the interaction, i.e. in average the $p_T$
shift tends to zero. Therefore, the above mechanism will not change the
results obtained \cite{3r} in DPM for soft collisions. \par

It turns out that in our formalism most of the observed suppression
takes place at very early times,
where the density of the medium is higher. Since hadron formation
times are longer, most of the
suppression takes place at a pre-hadronic (partonic) level.
Therefore, the mechanism described below is
not a conventional hadronic final state interaction and,
qualitatively, is expected to lead to similar
results as the jet quenching -- based on radiative parton energy
loss. Further discussion on this point
can be found in the conclusions.\\

\section{The Model}
\hspace*{\parindent}
The interaction of a large $p_T$ particle with
the soft medium is
described by the gain and loss differential equations which govern
final state interactions. In the following, the large $p_T$ particle
will be a $\pi^0$ and the medium will be all charged and neutral
secondaries produced in $AuAu$ collisions at $\sqrt{s_{NN}} = 200$~GeV.
Denoting by $\rho_H$ and $\rho_S$ the corresponding space-time
densities, we have following \cite{10r}

\beq \label{1e} {d\rho_H(x,p_T) \over d^4x} = - \widetilde{\sigma} \
\rho_S \left [ \rho_H (x, p_T) - \rho_H(x, p_T + \delta p_T)\right ]
\eeq

\noi where $\widetilde{\sigma}$ is the final state interaction
cross-section, averaged over the
momentum distribution of the colliding particles. The first term
describes the loss of $\pi^0$'s with a given $p_T$, due to its
interaction with the medium with density $\rho_S$. The second
term describes the (smaller) gain in the yield at a given $p_T$
resulting from the $\pi^0$'s produced at $p_T + \delta p_T$, which have
suffered a shift in $p_T$ due to the interaction. In the conventional
treatment \cite{10r} of eq. (\ref{1e}), one uses cylindrical space-time
variables with the longitudinal proper time $\tau = \sqrt{t^2 - z^2}$,
space-time rapidity $y = 1/2 \ell n ((t+z)/(t-z))$ -- to be identified
later on with the usual rapidity -- and transverse coordinate $s$. One
also assumes longitudinal boost invariance. Therefore, the above
picture is not valid in the fragmentation regions. One further assumes
that the dilution in time of the densities is only due to longitudinal
motion\footnote{Transverse expansion is neglected. The fact that HBT
radii are similar at SPS and RHIC and of the order of magnitude of the
nuclear radii, seems to indicate that this expansion is not large. The
effect of a small transverse expansion can presumably be taken into
account by a small change of the final state interaction
cross-section.} which leads to a $\tau^{-1}$ dependence on $\tau$.
\par

Eq. (\ref{1e}) can then be written in the form \cite{10r} \cite{11r} \beq
\label{2e} \tau \ {N_{\pi^0}(b,s,y, p_T) \over d \tau} = -
\widetilde{\sigma} N(b,s,y) \left [ N_{\pi^0}(b,s,y,p_T) - N_{\pi^0}(b,
s, y, p_T + \delta p_T) \right ] \eeq

\noi where $N(b,s,y) \equiv dN/dy$ $d^2s(y,b)$ is the density of all
charged plus neutral particles per unit rapidity and per unit of
transverse area at fixed impact parameter, integrated over $p_T$.
$N_{\pi^0}(b,s,y,p_T)$ is the same quantity for $\pi^0$'s at fixed
$p_T$. \par

Equation (\ref{2e}) has to be integrated from initial time $\tau_0$
to freeze-out time $\tau_f$. It is invariant under the change $\tau \to
c\tau$ and, thus, the result depends only on the ratio $\tau_f/\tau_0$. We
use the inverse proportionality between proper time and densities and
put $\tau_f/\tau_0 = N(b,s,y)/N_{pp}(y)$ where $N_{pp}(y) = (1 /\pi
R^2_p)$ $dN_{pp}/dy$ is the density of charged and neutral particles
per unit rapidity for minimum bias $pp$ collisions at $\sqrt{s} = 200$~
GeV. At $y^* \sim 0$, $N_{pp}(0) = 2.24$~fm$^{-2}$. This density is
about 90 \% larger than at SPS energies. Since the corresponding
increase in the $AA$ density is comparable, the average duration time
of the interaction will be approximately the same at CERN-SPS and RHIC,
about 5 to 7 fm. \par

Note that $N(b,s,y)$ in eq. (\ref{2e}) is the density at time $\tau_0$,
i.e. the density produced in the primary collisions. It can be computed
in DPM. The procedure is explained in detail in \cite{9r}. The hard
density $N_{\pi^0}$ in the primary collision is assumed to scale with
the number of binary collisions\footnote{Actually, due to the $p_T$
broadening, this scaling is not satisfied and its violation depends on
the value of $b$ (see column I ot Table 1). However, if we
incorporate this violation by changing $n(b,s)$ into $n(b,s) f(b)$ in
eq. (\ref{3e}), the value of $S_{\pi^0}$ will not change.}. \par

Eq. (\ref{2e}) can be easily integrated over $\tau$. We obtain in this
way the suppression factor $S_{\pi^0}(b,y,p_T)$ of the yield of
$\pi^0$'s at given $p_T$ and at each impact parameter, due to its
interaction with the dense medium. We get

\beq \label{3e} S_{\pi^0}(b,y, p_T) = {\int d^2s\ \sigma_{AB}(b) \ n(b,s)
\ \widetilde{S}_{\pi^0}(b,s,y,p_T) \over \int d^2s\ \sigma_{AB}(b)\
n(b,s)} \ , \eeq

\noi where the survival probability is given by

\beq \label{4e} \widetilde{S}_{\pi^0}(b,s,y,p_T)  = \exp \left \{ -
\widetilde{\sigma} \left [ 1 - {N_{\pi^0}(b,s,y,p_T + \delta p_T)\over
N_{\pi^0}(b,s,y,p_T)}\right ] N(b,s,y) \ell n \left ( {N(b,s,y) \over
N_{pp}(y)}\right ) \right \}\ . \eeq

\noi Here $\sigma_{AB}(b) = \{ 1 - \exp [-\sigma_{pp} AB \ T_{AB}(b)]\}$,
where $T_{AB}(b) = \int d^2s T_A(s) T_B(b-s)$, and $T_A(b)$ are profile
functions obtained from the Woods-Saxon nuclear densities \cite{12r}.
Upon integration over $b$ we obtain the $AB$ cross-section.
$n(b,s)$ is given by

\beq \label{5e} n(b,s) = AB \ \sigma_{pp}\ T_A(s) \ T_B(b-s)
/\sigma_{AB}(b) \ . \eeq

\noi Upon integration over $s$ we obtain the average number of binary
collisions at fixed $b$, $n(b)$. Note that if we neglect the second
term in eqs. (\ref{1e}) and (\ref{2e}), the factor inside brackets in
eq. (\ref{4e}) reduces to unity and we recover exactly the formula
\cite{9r} \cite{13r} for the survival probability of the $J/\psi$. \\

\section{Numerical Results}
\hspace*{\parindent}
In order to perform numerical
calculations, we need the value of
$\widetilde{\sigma}$ (which will be treated as a free parameter) as
well as the $p_T$ distribution of the $\pi^0$'s. Let us
concentrate first on $\pi^0$ production at mid-rapidities $(|y| < 0.35)$.\par

In $pp$ collisions at $\sqrt{s} = 200$~GeV, the shape of the $p_T$
distribution of $\pi^0$ can be described as $(1 + p_T /p_0)^{-n}$ with
$p_0 = 1.219$~GeV/c and $n = 9.99$ \cite{14r}. The corresponding
average $p_T$ is $<p_T> = 2p_0/(n-3) = 0.349$~GeV/c. The corresponding
value in central ($n_{part} = 350$) $AuAu$ collisions at the same
energy is $<p_T> = 0.453$~GeV/c. This value is obtained from ref.
\cite{15r} as an average of $\pi^+$ and $\pi^-$. \par

This is the well known $p_T$ broadening, which can be described as a
result of initial state interaction (see refs. [9b] \cite{16r} for the
case of $J/\psi$). Since it is not our purpose here to describe the
$p_T$ broadening we take it from experiment. Of course, the data also
contain the effect of the final state interaction. However the effect
of the latter
is only important at medium and large $p_T$ and it hardly changes
the value of $<p_T>$
(from the calculation in ref. [9b] this change is of the order of 1
\%). As discussed above, two mechanisms are responsible for the $p_T$
broadening. First, the decrease of the shadowing with increasing
$p_T$. This produces an increase of the ratio

\beq \label{6e} R_{AA}(b,y,p_T) = {dN^{AA} \over dy d^2p_T}(b)/n(b)
{dN^{pp} \over dyd^2p_T} \eeq

\noi from its small $p_T$ value (substantially lower than unity
\cite{3r}) to one. This contribution to the $p_T$ broadening can be
computed \cite{4r} at each value of $s$, $y$, $p_T$ and $b$ without
adjustable parameters in terms of the (experimentally known)
diffractive cross-section. The shadowing computed in this way describes
the EMC effect \cite{4r}. Moreover, incorporated in DPM, it gives a
good description of the centrality dependence of the charged particle
inclusive spectra in nucleus-nucleus collisions both at SPS and RHIC
energies \cite{3r}. The second mechanism is the Cronin effect proper,
which produces an increase of $R_{AA}$ above unity. It turns out that
at low $p_T$, where $R_{AA}$ is below unity, most of the increase of
$R_{AA}$ with $p_T$ is due to the first mechanism. However, the second
is not negligible and it is difficult to compute. In view of that we
proceed as follows. We assume that the $p_T$ distribution of $\pi^0$'s
in $AuAu$ at each $b$ is obtained from the corresponding one in $pp$
by keeping the same value of $n = 9.99$ and changing the scale $p_0$
into $p_0(b) = <p_T>_b$ $(n-3)/2$. In this way, the average  $p_T$ of 
the new distribution coincides with the measured value $<p_T>_b$ at 
the corresponding centrality \cite{15r}. We can
thus compute the ratio $R_{AA}$, eq. (\ref{1e}), in the absence of
final state interaction. The reasons for assuming that $n$ is not
changed are twofold. On the theoretical side, this is needed in order
to reproduce the $p_T$ broadening due to the variation of shadowing
with $p_T$. Indeed, at large $p_T$ the shadowing vanishes and the ratio
$R_{AA}$ is independent of $p_T$ -- which requires that $n$ is
constant\footnote{In perturbative QCD, $R_{AA}$ should tend to unity at large
$p_T$. However, this may occur at much larger values of $p_T$ than
present ones.}. On the experimental side, we notice that the value of
$n$ obtained in $dAu$ is the same as in $pp$ within errors. Note also
that a substantial change in $n$ would lead to a strong variation of
$R_{AuAu}$ and $R_{dAu}$ with $p_T$ at large $p_T$ -- which does not
seem to be the case experimentally.\par

In order to fix the absolute normalization we use the value of
$R_{AA}$ obtained from DPM in the case of soft collisions (i.e.
integrated over $p_T$) which describes well the experimental results at
all centralities \cite{3r}. In this way we obtain for the 10~\% most central
collisions ($n_{part} = 325$) the result shown in Table 1 (column I).
\par

To these values we apply the correction due to the suppression factor
$S_{\pi^0}$ in eq. (\ref{3e}). First, we neglect the second term in
eqs. (\ref{1e}) and (\ref{2e}). The formula is then identical to the
one for the $J/\psi$ case, as discussed above, and the suppression is
independent of $p_T$. In order to normalize our result to the
experimental values of $R_{AA}$ at large $p_T$, we use $\widetilde{\sigma} =
1.03$~mb, which gives a suppression factor $S_{\pi^0} = 0.143$. The
corresponding results are
given by column II in Table 1. We see that $R_{AA}$ increases
slightly with $p_T$ and agrees with experiment for $p_T > 5$~GeV/c. At
lower $p_T$ the result is significantly lower than the data. This is to
be expected. Indeed, as discussed above, the suppression
factor $S_{\pi^0}$ has to vanish at small $p_T$ -- which is not the
case so far. Before introducing this requirement, let us introduce the
second term in eqs. (\ref{1e}) and (\ref{2e}) and let us assume that
the $p_T$ shift of the $\pi^0$, due to its interaction with the medium,
is constant. Consider two cases~: $\delta p_T = 0.5$~GeV/c and $\delta
p_T = 1.5$~GeV/c. Imposing in all cases the same normalization at $p_T$
= 7~GeV/c \footnote{This is done by changing the only free parameter 
available ($\widetilde{\sigma}$ in eq. (\ref{4e})) in such a way that 
$\widetilde{\sigma}
[1 - 
N_{\pi^0}(b,s,y,p_T + \delta p_T)/N_{\pi^0}(b,s,y,p_T)]=1.03$. This is 
done at $p_T = 7$~GeV/c, $\eta^* = 0$ for the 10~\% most central 
collisions. Of course, the same value of $\widetilde{\sigma}$ is then 
used for all values of $p_T$, $\eta$ and $b$.} we obtain the results 
in columns III and IV of Table 1,
respectively and in Fig.~1. We observe a slight increase of $R_{AA}$ 
at large $p_T$,
consistent with the data for $p_T > 7$~GeV/c. The results tend to those in
column I with increasing value of $\delta p_T$ -- as it should be. The
important result here is that, with constant $\delta p_T$, one obtains
a small increase of $R_{AA}$ with $p_T$ consistent with the data at
large $p_T$ ($p_T \ \gsim\ 5$~GeV/c). The result is rather insensitive
to the value of the shift, for any $\delta p_T \ \gsim \
0.5$~GeV/c. Of course, the problem at small $p_T$ remains. In order to
cure it, we assume that $\delta p_T \propto (p_T - <p_T>_b)$. In this
case the factor $S_{\pi^0}$ is 1 at $p_T = <p_T>$ as it should be. Taking
$\delta p_T = (p_T - <p_T>_b)/20$ we obtain the values in column V
of Table 1 and in Fig.~1. Note that these values are rather insensitive to the
fraction (5 \%) of $p_T$ lost in each interaction. Varying this
fraction between 1~\% and 10~\% the results do not change
substantially. We see from Table 1 and Fig.~1 that the slight 
increase of $R_{AA}$
obtained with constant $\delta p_T$ is changed into a slight
decrease, also consistent with
the data, and, moreover, the agreement at small $p_T$ is significantly
improved. Actually, a better agreement in the low $p_T$ region is
obtained assuming

\beq
\label{7e}
\delta p_T = \left ( p_T \ - \ <p_T>_b\right )^{1.5}/20 \ .
\eeq

\noi  The results are given in column VI of Table 1 and in Fig. 2.\par

As discussed above, we assume $\delta p_T \propto (p_T - 
<p_T>)^{\alpha}$ in order to implement the condition $S_{\pi^0} \to 
1$ as
$p_T \to <p_T>$. The power $\alpha$ controls the way in which this 
limit is reached.
Although such a behaviour is needed at low $p_T$, it does not have to be
the same at large $p_T$. (In fact a power larger than unity cannot be
used at large $p_T$ since $\delta p_T$ would be larger than $p_T$).
Actually one obtains an equally good agreement
with data using $\delta p_T = (p_T - <p_T>_b)^{1.5}/20$ for $p_T
< 7$~GeV/c and $\delta p_T = (7 - <p_T>_b)^{1.5}/20$ for $p > 7$~GeV/c. 
In this
case the result is given in column VII of Table 1 and in Fig.~2. \par

The important result is that at large $p_T$
($p_T \ \gsim\ 5$~GeV/c)
we obtain a slight increase of $R_{AA}$ with $p_T$ for constant $\delta
p_T$ and a slight decrease for $\delta p_T \propto p_T$ (or
$p_T^{1.5}$). In both cases there is agreement with PHENIX data [1a].\par

The centrality dependence of $R_{AA}(p_T)$ at large $p_T$ ($p_T >
4$~GeV/c) is reasonably well described (see Figs.~2 and 3). The constancy of
$R_{AA}(p_T)$ ($p_T >
4$~GeV/c) for $N_{part} < 60$ is the result of a cancellation, in this range
of $N_{part}$, between the increase of the $p_T$ broadening and the increase of
the suppression with increasing centrality. 
This
centrality dependence has been reproduced in a recent work,
also based on absorption in a dense medium \cite{17r}. \par

We turn next to minimum-bias $dAu$ collisions. Here $<p_T> =
0.39$~GeV/c \cite{18r}. With the same value of $n$ as above $(n
=9.99)$, this corresponds to $p_0 = 1.346$. Calculating the ratio
$dAu$ to $pp$ and fixing the normalization from the DPM value
(integrated over $p_T$) of this ratio, we obtain the result in Table 2.
These values, obtained without introducing nuclear absorption, are in 
agreement with experiment in the lower half of the $p_T$ range. At 
large
$p_T$, nuclear absorption is expected to be present both in $dAu$ and
$AuAu$ collisions. The $dAu$ data at large $p_T$ are consistent with
the presence of nuclear absorption. However, the error bars are too
large in order to perform a quantitative study of this question -- and
determine the value of $\sigma_{abs}$\footnote{Introducing nuclear
absorption in $AuAu$ collisions would result in a smaller value of
$\widetilde{\sigma}$.}.

Finally, we turn to the $dAu$ collisions at forward rapidities.
Consider first the ratio $R_{dAu}$ integrated over $p_T$. In our
approach, $R_{dAu}$ decreases as $y^*$ increases. There are two
effects contributing to this decrease. \par

The first effect is basically due to
energy-momentum conservation. It has been known for a long time in
hadron-nucleus collisions at SPS energies (as well as at lower ones)
and it is well understood in string models such as DPM and QGSM.
Recently, it has been referred to as the low $p_T$ ``triangle''
\cite{19r}. Its extreme form occurs in the hadron fragmentation region,
where the yield of secondaries in collisions off a heavy nucleus is
smaller than the corresponding yield in hadron-proton. This phenomenon
is known as nuclear attenuation. It turns out, that, at RHIC energies,
this effect produces a decrease of $R_{dAu}$ of about 30~\% between
$\eta^* = 0$ and $\eta^* = 3.2$. \par

The second effect is the increase of the shadowing corrections in $dAu$
with increasing $y^*$ \cite{4r}. This produces a decrease of
$R_{dAu}(p_T)$ between $\eta^* = 0$ and $\eta^* = 3.2$ of about 30~\%
for pions produced in minimum bias collisions\footnote{The situation is
different in $AuAu$. Here the shadowing decreases with increasing
$\eta$, but its variation is much smaller than in $dAu$ \cite{4r}.}. This
decrease is practically independent of $p_T$. Therefore, we expect a
suppression factor of about 1.7 between $R_{dAu}(p_T)$ at $\eta^* = 0$
and at $\eta^* = 3.2$, practically independent of $p_T$. This is 
consistent with the BRAHMS results [2b].\par

The same result is obtained with the procedure used above in $AuAu$
collisions, if the $<p_T>$ of pions in $dAu$ is the same at $\eta^* =
0$ and at $\eta^* = 3.2$. This is approximately the case in $AuAu$
collisions [1b] and, in our approach, it is expected also in $dAu$.
Indeed, as discussed above, most of the $p_T$ broadening at low $p_T$
is due to the variation of shadowing with $p_T$ -- and this variation
is practically the same at $\eta^* = 0$ and at $\eta^* = 3.2$. On the
contrary, the dependence of shadowing on centrality at fixed $\eta$ is
quite important and, thus, the centrality dependence of 
$R_{dAu}(p_T)$ is quite large. Details with be presented 
elsewhere\footnote{After completion of this work, we learned of a 
related work, ref. \cite{20r}.
In that
paper only Cronin effect plus geometrical shadowing was considered in 
$dAu$ collisions. So the authors are
unable to describe data at large $\eta$ (contrary to our approach, where
dynamical shadowing is taken into account).}.\par

\section{Conclusions}
\hspace*{\parindent} In this work the suppression of $\pi^0$ 
production at large $p_T$ in $AuAu$ collisions is described in 
terms of final state interaction in the dense medium produced in the 
collision. The mechanism is similar to the one responsible for 
$J/\psi$ suppression. 
A 
nice feature of our formulation is that it leads to a suppression of 
$R_{AA}(p_T)$ at large $p_T$ ($p_T > 5$~GeV/c) which is rather 
insensitive to the size and form of the $p_T$ shift produced by the 
final state interaction. \par

Our approach contains dynamical, non-linear, shadowing. This 
shadowing is determined in terms of (experimentally known) 
diffractive cross-sections. As $s \to \infty$, it leads to saturation 
of the parton distributions. However, at both RHIC and LHC energies, 
this shadowing is significantly smaller than the one present in a 
saturation regime. By itself, the shadowing in our approach is not 
sufficient to explain the difference in $R_{dAu}$ between $\eta^* = 
0$ and $\eta^* = 3.2$ measured by BRAHMS. Indeed, it produces only 
about one half of the measured variation. However, when the effect of 
shadowing (i.e. the increase of shadowing with increasing rapidity) 
is combined with low $p_T$ effects present in string models such as 
DPM and QGSM, agreement with the BRAHMS measurement is achieved. This 
also shows that decreasing $x$ by increasing energy is not equivalent 
to doing so by going to forward rapidities. In the latter case the 
value of $R_{dAu}$ at low $p_T$ is substantially reduced -- which 
obviously influences its large $p_T$ value. This is not so in the 
former case. \par

In this paper we have restricted ourselves to the study of the
$\pi^0$ large $p_T$ suppression. Other
observables have been measured such as the back-to-back and near side
large $p_T$ azimuthal
correlations and large $p_T$ azimuthal anisotropy. As discussed in
\cite{21r}, the present data on these
observables indicate that the suppression takes place at very early
times, where the density of the
medium is larger. However, this does not imply that the suppression
is due to non-abelian 
radiative parton energy loss.
Indeed, as discussed in Section 1, in
our mechanism the suppression also
takes place at very early times, at a partonic level. In view of
that, we expect that the results for
these observables will be similar to the ones obtained in the radiative
jet
quenching scenario. In our approach the
back-to-back correlation will be suppressed by the same amount as the
single particle one, whereas the
near-side one will remain essentially the same -- since the two large
$p_T$ particles originate from the
same jet. Likewise, a high $p_T$ azimuthal anisotropy is expected in
our approach due to the asymetry of
the dense medium at early times.\par

The large $p_T$ suppression phenomenon is important in order to
determine the properties of the dense
medium in which it takes place. In our approach, using the inverse
proportionality between density and
interaction time, we find that in a central $Au$ $Au$ collisions the
density of the medium is about 5
times larger than in $pp$. This factor is predicted to be practically
unchanged at LHC energies. In a
radiative
jet quenching scenario much larger densities have been claimed in the
literature. However, in a recent
paper \cite{22r}, the density of the medium has been found to increase by a
factor 4 between peripheral and
central $Au$ $Au$ collisions at RHIC -- consistent with our result. \par

In order to distinguish between radiative
jet quenching scenario and
collisional energy loss or collisional
$p_T$-shift it will be very interesting to measure the detailed
medium modifications of jet shapes and
multiplicities \cite{23r}. While we do not see at present how to
disentangle the different scenarios on the
basis of qualitative arguments, there will hopefully be differences
in their predictions at a
qualitative level.\\

\noi {\Large \bf Acknowledgments} \par\nobreak

It is a pleasure to thank N. Armesto, A. Krzywicki, C. Pajares,
C. Salgado and D.
Schiff for interesting discussions. We also thank D. d'Enterria for
discussions and information on the PHENIX data. Two of the authors
(A.C. and A. B. K. ) acknowledge final support by INTAS grant 00-00366.
This work was also supported by grants
RFBR 01-02-17383 and SSch-1774-2003.2  and by the Federal Program of the
Russian Ministry of Industry, Science and Technology 40.052.1.1.1112.

\newpage
\centerline{\bf CAPTIONS}
\vskip 1 truecm

\noi \underbar{\bf Table 1.} Values of $R_{AuAu}^{\pi^0}(p_T)$ for the
10~\% most central collisions $AuAu$ collisions at mid-rapidities
($|y^*| < 0.35$). Column I is the result obtained with no final state
interaction. The results in the other columns include final state
interaction with several ansatzs for the $p_T$ shift induced by this
interaction (see main text).\\

\noi \underbar{\bf Table 2.} Values of $R_{dAu}^{\pi^0}(p_T)$ for
minimum bias $dAu$ collisions at mid rapidities ($|y^*| < 0.35$).\\

\noi \underbar{\bf Fig 1.} Values of $R_{AuAu}^{\pi^0}(p_T)$ for the
10~\% most central collisions at mid-rapidities ($|y^*| <
0.35$), using the $p_T$ shift given by eq. (\ref{7e})
$\delta p_T = (p_T - <p_T>_b)^{1.5}/20$
(solide line), the linear case
$\delta p_T = (p_T - <p_T>_b)/20$
(dashed-dotted line), the cuadratic case
$\delta p_T = (p_T - <p_T>_b)^2/20$ (dashed line),
and $\delta p_T =$ constant (dotted lines). See Table 1 and main text for
details.\\

\noi \underbar{\bf Fig 2.} Values of $R_{AuAu}^{\pi^0}(p_T)$ for the
10~\% most central collisions (lower line) and for peripheral
(80-92~\%) collisions (upper line) at mid-rapidities ($|y^*| <
0.35$), using the $p_T$ shift given by eq. (\ref{7e}),
$\delta p_T = (p_T - <p_T>_b)^{1.5}/20$,
  (solide line).
The dashed line is obtained using eq.~(\ref{7e}) for $p_T  \leq
7$~GeV/c and $p_T =$ constant for $p_T \geq 7$~GeV/c (see main text).
The data are from ref. [1a].\\

\noi \underbar{\bf Fig 3.} Centrality dependence of $R_{AuAu}^{\pi^0}$
for $p_T > 4$~GeV/c using the $p_T$ shift given by eq. (\ref{7e}).
The data are from ref. [1a].\\

\newpage
\begin{center}
{\bf TABLE 1}\par \vskip 1 truecm

\begin{tabular}{|c|l|l|l|l|l|l|l|}
\hline
$p_T$ (GeV/c) &I &II &III &IV &V &VI &VII\\
\hline
0.5 &0.38 &0.05 &     &     &0.34 &0.38 &0.38\\
2   &0.90 &0.13 &0.08 &0.11 &0.20 &0.31 &0.31 \\
5   &1.48 &0.21 &0.18 &0.19 &0.23 &0.25 &0.25 \\
7   &1.69 &0.24 &0.24 &0.24 &0.24 &0.24 &0.24\\
10  &1.84 &0.27 &0.34 &0.30 &0.25 &0.24 &0.32 \\
\hline
\end{tabular}
\end{center}

\par \vskip 2 truecm

\begin{center}
{\bf TABLE 2}\par \vskip 1 truecm

\begin{tabular}{|c|c|}
\hline
$p_T$ (GeV/c) &$R_{dAu}(p_T)$\\
\hline
0.39 &0.63 \\
1 &0.78 \\
2 &0.92 \\
3 &1.01\\
5 &1.10 \\
7 &1.16 \\
10 &1.21 \\
\hline
\end{tabular}
\end{center}

\newpage

\centerline{\bf Figure 1}
\vspace{1cm}

\begin{center}
\hspace{-1.2cm}\epsfig{file=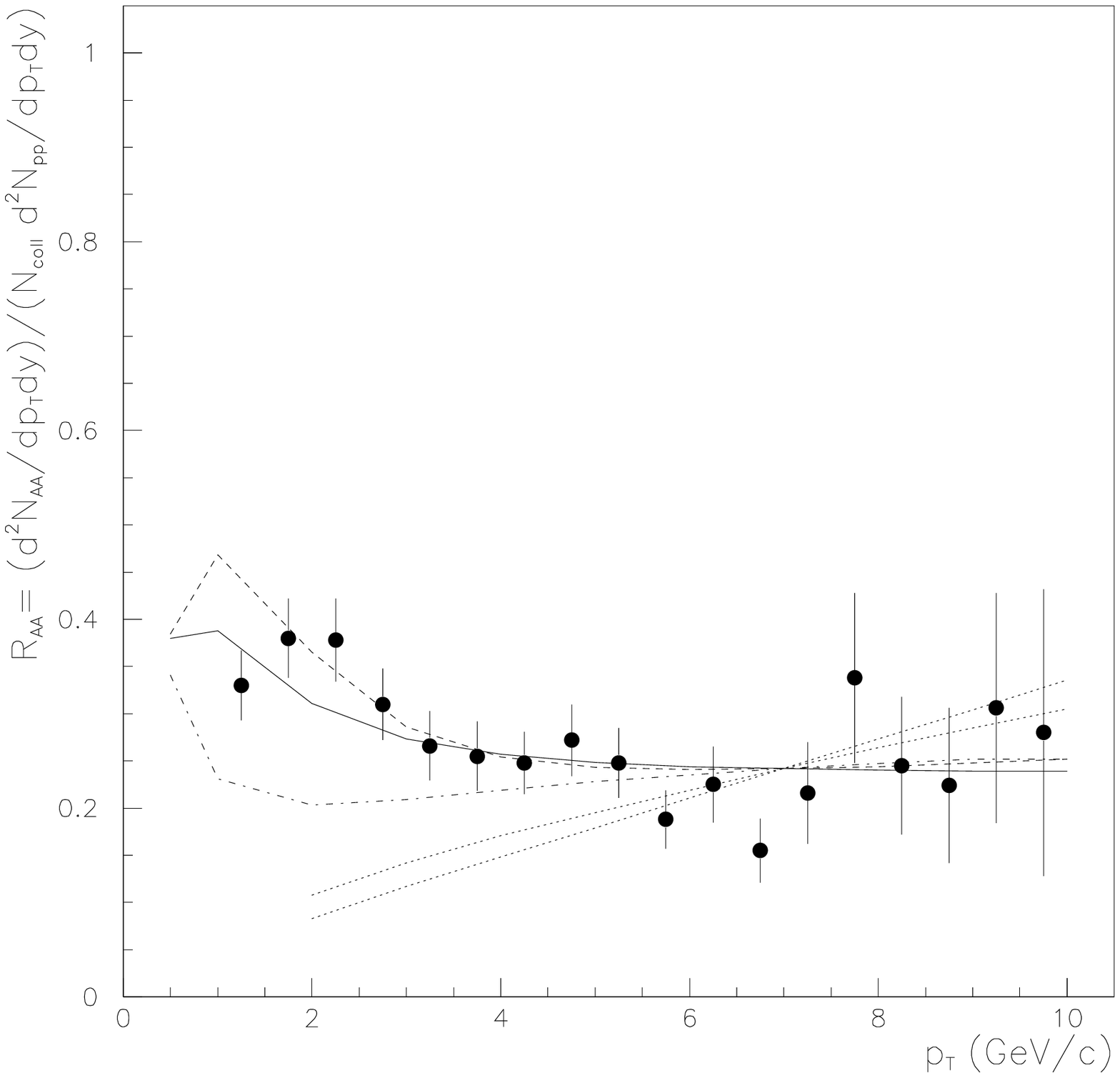,width=17.cm}
\end{center}

\newpage

\centerline{\bf Figure 2}
\vspace{1cm}

\begin{center}
\hspace{-1.2cm}\epsfig{file=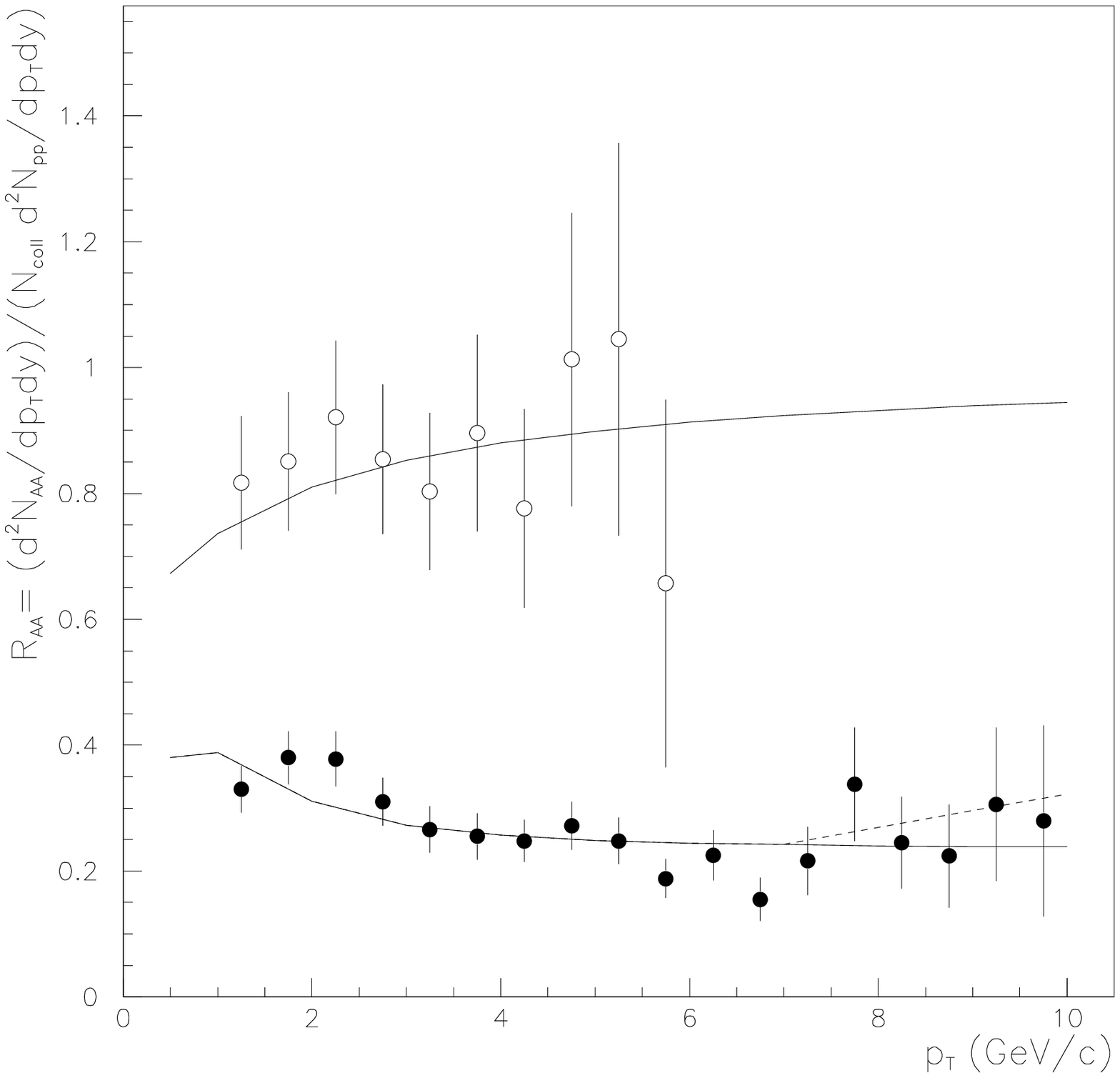,width=17.cm}
\end{center}

\newpage

\centerline{\bf Figure 3}
\vspace{1cm}

\begin{center}
\hspace{-1.2cm}\epsfig{file=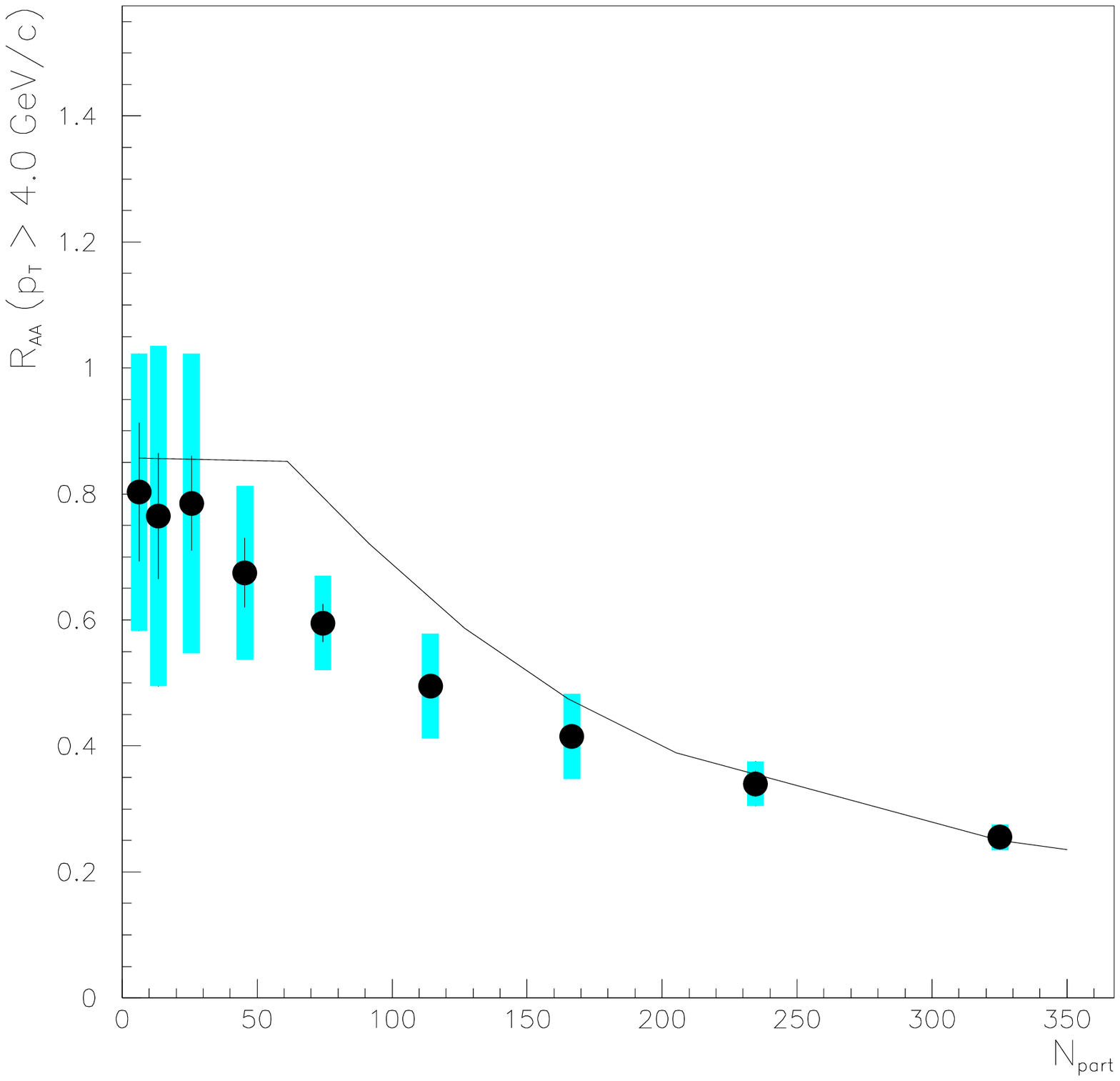,width=17.cm}
\end{center}

\end{document}